\documentclass{WileyMSP-template}
\usepackage{siunitx}
\usepackage{booktabs}
\usepackage{amsmath}
\usepackage{minted}
\usepackage{hyperref}
\hypersetup{colorlinks = true, allcolors = .}
\usepackage{enumitem}
\usepackage{ragged2e}
\justifying

\usepackage{tikz}
\usetikzlibrary{shapes,arrows,positioning,fit,backgrounds,decorations.pathmorphing,decorations.pathreplacing,calc,patterns,arrows.meta}
\usepackage{pgfplots}
\pgfplotsset{compat=1.18}
\usepackage{smartdiagram}

\begin{document}

\pagestyle{fancy}

\title{Towards knowledge-based workflows: a semantic approach to atomistic simulations for mechanical and thermodynamic properties}
\maketitle


\author{Abril Az\'{o}car Guzm\'{a}n *}
\author{Hoang-Thien Luu}
\author{Sarath Menon}
\author{Tilmann Hickel}
\author{Nina Merkert}
\author{Stefan Sandfeld}

\begin{affiliations}
Dr. A. A. Guzm\'{a}n, Prof. Dr. S. Sandfeld\\
Institute for Advanced Simulations – Materials Data Science and Informatics (IAS‑9), Forschungszentrum J\"{u}lich GmbH, 52425 J\"{u}lich, Germany\\
* a.azocar.guzman@fz-juelich.de

Dr. H.-T. Luu, Prof. Dr. N. Merkert\\
Institute of Metallurgy - Chair of Computational Material Sciences/Engineering, Technical University of Clausthal, 38678 Clausthal-Zellerfeld, Germany

Dr. S. Menon\\
Interdisciplinary Centre for Advanced Materials Simulation (ICAMS), Ruhr-Universit\"at Bochum, 44801 Bochum, Germany.

Dr. T. Hickel\\
Bundesanstalt für Materialforschung und -pr\"{u}fung, 12489 Berlin, Germany

\end{affiliations}



\begin{abstract}
Mechanical and thermodynamic properties, including the influence of crystal defects, are critical for evaluating materials in engineering applications. Molecular dynamics simulations provide valuable insight into these mechanisms at the atomic scale. However, current practice often relies on fragmented scripts with inconsistent metadata and limited provenance, which hinders reproducibility, interoperability, and reuse. FAIR data principles and workflow-based approaches offer a path to address these limitations. We present reusable atomistic workflows that incorporate metadata annotation aligned with application ontologies, enabling automatic provenance capture and FAIR-compliant data outputs. The workflows cover key mechanical and thermodynamic quantities, including equation of state, elastic tensors, mechanical loading, thermal properties, defect formation energies, and nanoindentation. We demonstrate validation of structure–property relations such as the Hall–Petch effect and show that the workflows can be reused across different interatomic potentials and materials within a coherent semantic framework. The approach provides AI-ready simulation data, supports emerging agentic AI workflows, and establishes a generalizable blueprint for knowledge-based mechanical and thermodynamic simulations.
\end{abstract}

\section{Introduction}
\label{Sec:intro}

Atomistic modelling provides a foundation for understanding mechanical properties, defect dynamics, and phase-transformation processes in materials. Like other computational studies, it was historically conducted through ad-hoc scripts that combined method choices, parameters, and analysis steps in a non-standardized manner. The shift toward systematic workflows and high-throughput infrastructures, such as the Materials Project \cite{Jain2013, Horton2025}, AFLOW \cite{CURTAROLO2012}, NOMAD repository \cite{Scheidgen2023}, has shown the strong demand for scalable and reproducible computational methodologies. Workflow-based automation has also proven valuable for developing and validating machine-learning interatomic potentials (MLIPs), for example through frameworks that support data generation, potential fitting, and large-scale validation \cite{Menon2024}. As density functional theory (DFT), empirical potentials, and MLIPs continue to expand the scope and fidelity of simulations, the resulting data landscape has grown in both volume and complexity, creating a pressing need for structured and interoperable approaches \cite{Himanen2019}.

Based on these developments, molecular dynamics (MD) techniques become increasingly powerful. Despite significant advances, however, they continue to face challenges related to non-standardized workflows and data representations, as outlined above. Most studies rely on fragmented, script-based simulations in which input parameters, analysis steps, and methodological assumptions are kept within monolithic input files. This lack of standardization in metadata and provenance makes it difficult to reproduce simulation protocols across research groups or even within the same project over time. Reuse of workflows is often limited, particularly when switching between modelling approaches that use different software, each having its own conventions and file formats. Moreover, simulation outputs are rarely integrated into broader research data infrastructures, which restricts their accessibility and reuse. API-based initiatives such as OPTIMADE \cite{Evans2024} and PLUMED \cite{Bonomi2019} help promote transparency and reproducibility in atomistic simulations data, but do not address the deeper semantic and interoperability challenges discussed here.

Therefore, a transition toward FAIR research data is essential for advancing computational materials science. The FAIR principles require that data be findable, accessible, interoperable, and reusable, which in turn demands standardized metadata and machine-readable provenance \cite{Wilkinson2016}. Specific to the materials community, FAIR data practices can enable new research directions and require shared, domain-aware metadata \cite{Scheffler2022, Ghiringhelli2023}. Semantic interoperability is particularly important for enabling cross-study comparison, combining results from different simulation methods, and integrating computational data with experimental measurements. Materials science ontologies have proven effective for annotating metadata while capturing complex descriptions of materials, methods, and workflows in a consistent manner \cite{Bayerlein2022, Bayerlein2024}. These semantic representations provide the foundation for knowledge graphs and artificial intelligence (AI) workflows that can search, interpret, and reason over large bodies of simulation data.

Emerging AI technologies increasingly rely on structured knowledge representations to support reasoning and decision making \cite{Ghafarollahi2024}. Materials science knowledge graphs provide these capabilities by encoding relationships among materials, simulation methods, and studied properties in a machine-readable format. This structure allows downstream tasks such as complex querying, verification of structure–property relationships, automated selection of modeling strategies, and rapid exploration of high-dimensional data landscapes. As a result, workflows enriched with semantic metadata are well positioned to serve as the backbone of next-generation AI-assisted materials research.

Although several workflow engines exist for atomistic and in particular DFT simulations \cite{Janssen2019, Rosen2024, Huber2020}, most lack the semantic rigor or domain specificity needed to support interoperable and reusable workflows. Current standardization efforts largely focus on electronic-structure data and rarely extend to the calculation of mechanical properties, thermodynamic quantities, or crystal defect energetics that are essential for engineering applications. As a result, no unified framework is available that semantically captures both mechanical behavior and thermodynamic processes across different simulation methods.

\begin{figure}[htbp]
    \centering
    \includegraphics[width=0.65\textwidth]{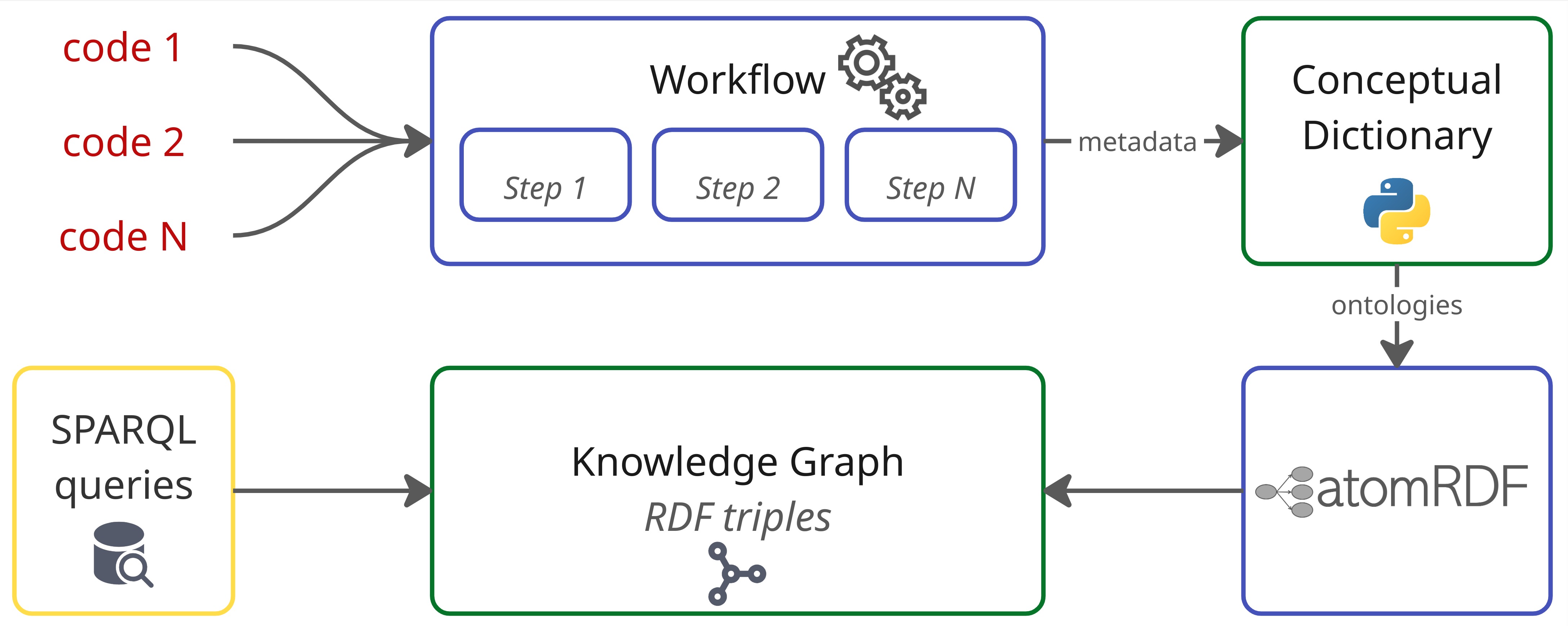}
    \caption{Data journey proposed for atomistic simulations, representing data resulting from atomistic software codes into a knowledge graph.}
    \label{fig:data-journey}
\end{figure}

We propose semantically annotated workflows for atomistic simulations of mechanical and thermodynamic properties. The data journey for the proposed framework is schematized in Figure \ref{fig:data-journey}. It is specifically intended for calculations of the equation of state, elastic tensor, compression and tensile loading, nanoindentation, thermal expansion coefficient specific heat, Gibbs free energies, and defect formation energies. We demonstrate the reusability of the workflows across different empirical potentials, including EAM, MLIPs and universal MLIPs. We also show targeted queries for physical insight, such as the validation of Hall-Petch strengthening law. We provide a generalizable blueprint to produce FAIR and AI-ready data from atomistic simulations to be integrated with research data infrastructures and knowledge graphs.

\section{Methodology}

\subsection{Workflow architecture} \label{sec:workflow_env}

\begin{figure}[htbp]
    \centering
    \includegraphics[width=0.99\textwidth]{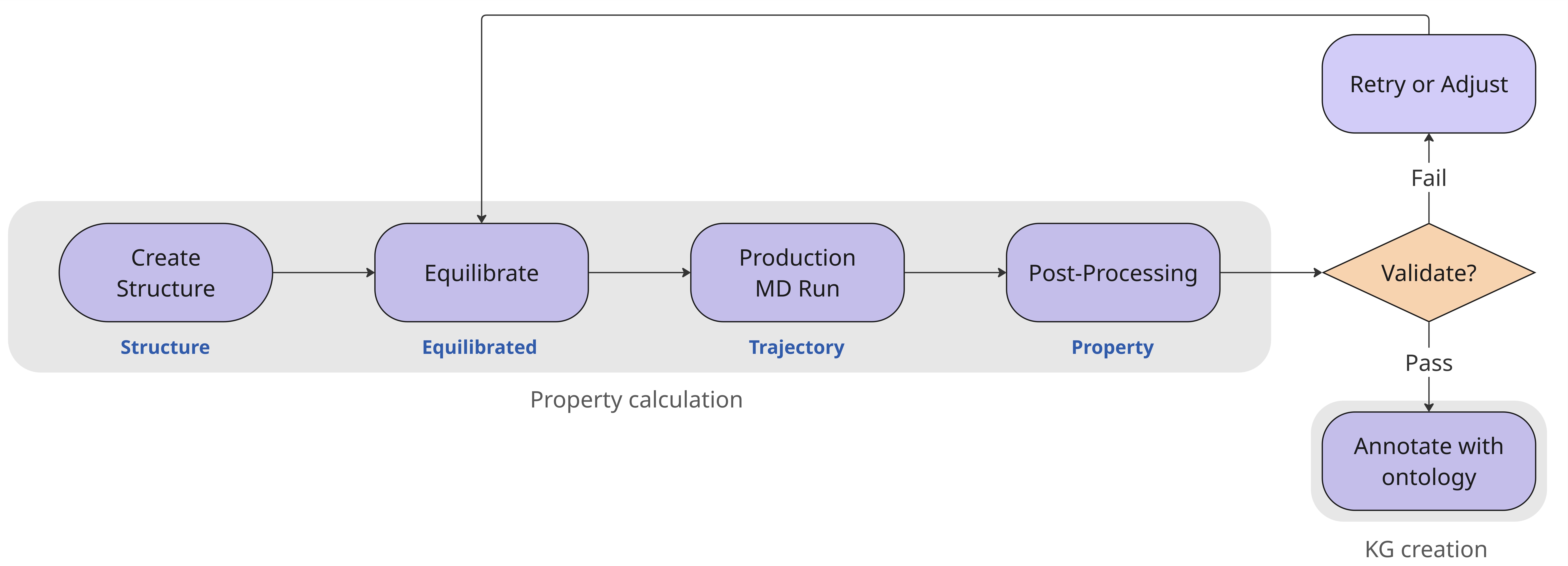}
    \caption{Workflow diagram showing the general sequence of steps used in all simulations, from structure creation to validation. Successful results are annotated with ontologies and added to the knowledge graph. The `KG creation' workflow is described in Section \ref{sec:semantic}, while the `Property calculation' workflows are described in further detail in Section \ref{sec:workflows}.}
    \label{fig:workflow_dag}
\end{figure}

Our workflow architecture follows a general sequence of steps, as shown in Fig.~\ref{fig:workflow_dag}. It consists of two stages: (i) a property calculation stage and (ii) knowledge graph (KG) creation. The first stage comprises the creation or import of an atomic structure, system equilibration, a production molecular dynamics simulation, and post-processing. Each step is implemented as an independent workflow node with clearly defined inputs and outputs, as well as an explicit execution order. This is followed by the KG creation stage, in which, after validation, the data and metadata generated in the previous stage are semantically annotated and integrated into the knowledge graph.

To ensure complete reproducibility in the spirit of Sec.~\ref{Sec:intro}, for the property calculation workflows, we employ a workflow management system that formalizes the scientific process into discrete, traceable units. Each step of the workflow is defined at a coarse-grained level, corresponding to a distinct scientific task, and represented as an independent workflow node with clearly specified inputs and outputs. Many existing systems, such as pyiron \cite{Janssen2019}, jobflow \cite{Rosen2024}, or file-based frameworks like the Common Workflow Language (CWL) \cite{Crusoe2022}, adopt a similar structural paradigm.\par\vspace{1em}

We use the pyiron software \cite{Janssen2019}, in a graph and node based implementation \cite{pyiron_workflow}, to construct and execute property calculation workflows. Our choice is motivated by its ability to directly handle complex Python objects without requiring file-based serialization. Beyond defining the core Python function, the only additional requirement is the use of a decorator, which enables the composition of multiple nodes into a complete and executable workflow. 
\par\vspace{1em}

In addition to the functionality provided by pyiron, we design the individual workflow nodes to be context-aware: each node generates a Python dictionary containing comprehensive scientific metadata for its respective step. These metadata records are subsequently aggregated and used to construct a knowledge graph in the KG creation stage, providing a structured and machine-interpretable representation of the workflow and its execution context, as described in more detail in Sec.~\ref{sec:semantic}.

An example is shown in Listing~\ref{list:cb1}, which illustrates the complete workflow for calculating the equation of state of a given system using a specified interatomic potential. In this case, the property calculation stage includes generating the initial structure and performing MD simulations with LAMMPS at a range of volumes, with the output of each step passed directly to the next. The metadata from each step is aggregated, and then parsed into the knowledge graph in the KG creation stage.

\begin{listing}[H]
\begin{minted}[linenos, fontsize=\small]{python}
from workflows.pyiron.build import bulk
from workflows.pyiron.evcurves import calculate_ev_curves
from pyiron_workflow import Workflow, as_function_node
from conceptual_dictionary import ConceptualDict 

pair_style = "eam/alloy"
pair_coeff = "* * workflows/potentials/Fe_Ack.eam Fe"

#ontology-aware metadata aggregation
cd = ConceptualDict()

#property calculation workflow
wf = Workflow('ev1')
wf.structure = bulk('Fe', cubic=True, cdict=cd)
wf.ev_curves = calculate_ev_curves(wf.structure, 
                                  pair_style, pair_coeff, cdict=cd)
wf.run()
cd.to_yaml("metadata.yaml")

#kg creation
from atomrdf import WorkflowParser, KnowledgeGraph

kg = KnowledgeGraph()
wf = WorkflowParser(kg)
wf.parse("metadata.yaml")

\end{minted}
\caption{Complete workflow for the calculation of equation of state. In the property calculation stage, an atomic structure is created, and the equation of state is calculated. Metadata is aggregated in each step, and then added to the knowledge graph.}  
\label{list:cb1}
\end{listing}

In this work, we provide reusable and reproducible workflows for the calculation of several material properties, including the equation of state, elastic constants (shear modulus, bulk modulus, and Poisson ratio), thermal expansion coefficient, and thermodynamic quantities such as specific heat and free energy, which can be used to determine phase transition temperatures. We also include workflows for mechanical tests under compression and tension. In addition, auxiliary nodes for generating atomic and polycrystalline structures are provided. The scientific methodology behind each of these nodes is briefly described in Sec. \ref{sec:workflows}, along with a discussion of representative results in Sec. \ref{sec:workflow_results}.
\par\vspace{1em}

\subsection{Semantic annotation and knowledge graph creation}\label{sec:semantic}

Compliance with the FAIR principles \cite{Wilkinson2016} requires describing data with metadata. Specifically, in terms of interoperability the metadata should use a formal, accessible, shared and broadly applicable language for knowledge representation, for which the standard choices are RDF or OWL \cite{w3c-rdf11,w3c-owl2}. These formal requirements are addressed at the community level through initiatives such as NFDI-MatWerk \cite{eberl_2021}, whose objective is to support the development and harmonization of semantic standards for interoperable materials research data. The domain-specific semantic artifacts used in this work are developed within this context. Beyond adhering to the FAIR principles, semantic interoperability in materials science presents the benefit of expressing structure-property relationships and in this way injecting physical knowledge to the data itself. 

\begin{figure}[htbp]
    \centering
    \includegraphics[width=0.99\textwidth]{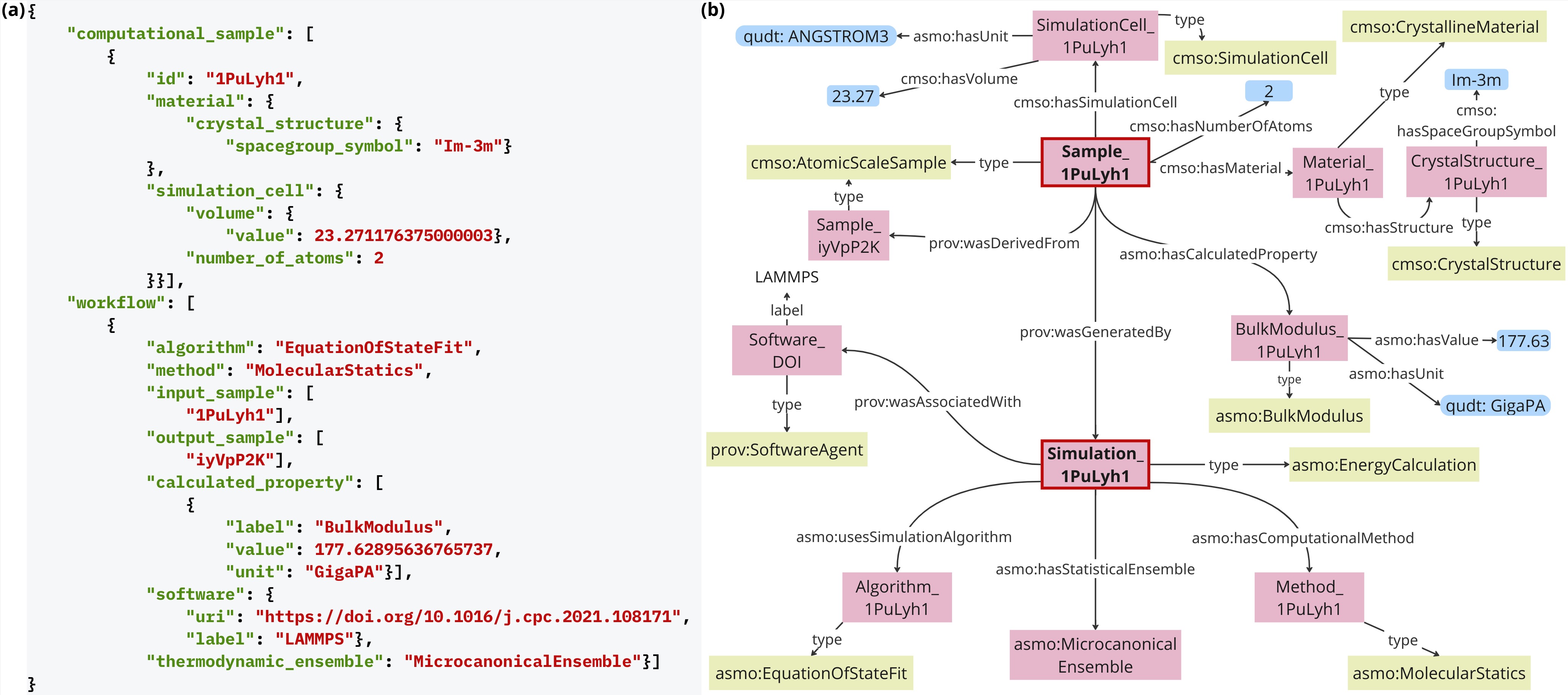}
    \caption{(a) Conceptual dictionary snippet in JSON format. (b) Schematic representation of the RDF serialization of the metadata fields shown in the JSON. The identifiers are shortened for visualization purposes, DOIs or unique identifiers are assigned to all instances. }
    \label{fig:metadata}
\end{figure}

To describe the metadata of the data produced in this paper we use the Computational Materials Sample Ontology (CMSO) \cite{CMSO} and the Atomistic Simulation Methods Ontology (ASMO) \cite{ASMO}. Concepts related to the computational sample, including the material specification, crystal structure, and simulation box are contained in CMSO. While ASMO describes the methods used for modelling, in this case molecular dynamics, the interatomic potential, and the simulation algorithms. The workflow aspects are represented using the provenance ontology PROV-O, a lightweight W3C standard ontology that defines a set of concepts for describing provenance information across different application domains \cite{w3c-provo}. As shown in Figure \ref{fig:data-journey}, the metadata of each workflow run is stored in a conceptual dictionary, a Python dictionary with keys aligned to the necessary ontology classes. In the example shown in Figure \ref{fig:metadata}(a), the \texttt{ConceptualDict} object aggregates the metadata generated at each step, enabling subsequent semantic integration. The JSON or YAML representation of the dictionary is then serialized into RDF triples using the atomRDF software \cite{atomRDF}, a schematic representation is shown in Figure \ref{fig:metadata}(b). The main objects are the computational sample and the simulation instance.

A key design principle of the \texttt{ConceptualDict} is its independence from any specific workflow engine. It provides a lightweight abstraction for capturing semantically relevant metadata without constraining workflow definition or execution. As a result, the same metadata structure can be populated across different workflow environments, ensuring consistent description of workflow execution and results. Listing~\ref{list:cb1job} shows the same workflow as Listing~\ref{list:cb1} implemented using jobflow, illustrating the environment-agnostic nature of the semantic annotation approach.

\begin{listing}[H]
\begin{minted}[linenos, fontsize=\small]{python}
from workflows.jobflow.build import bulk
from workflows.jobflow.evcurves import calculate_ev_curves
from jobflow import job, run_locally, Flow
from conceptual_dictionary import ConceptualDict 

pair_style = "eam/alloy"
pair_coeff = "* * workflows/potentials/Fe_Ack.eam Fe"

#ontology-aware metadata aggregation
cd = ConceptualDict()

#property calculation workflow
structure = bulk(name ='Fe', cubic=True, cdict=cd)
ev_curves = calculate_ev_curves(structure.output, 
                                   pair_style, pair_coeff, cdict=cd)
flow = Flow([structure, ev_curves])
run_locally(flow)

\end{minted}
\caption{Workflow for the calculation of equation of state using jobflow. Only the property calculation stage is shown as KG creation stage is identical to Listing \ref{list:cb1}.}  
\label{list:cb1job}
\end{listing}

\subsection{Property calculation workflows} \label{sec:workflows}

As discussed in Sec. \ref{sec:workflow_env} and \ref{sec:semantic}, we present workflows for predicting material properties from atomistic simulations that are scalable, reproducible, and both technically and semantically interoperable.
In the remainder of this section, we briefly describe the scientific background behind the property calculation workflows. The complete set of workflows, including all input parameters, is available in the corresponding repository \cite{repo-workflows}.

All atomistic simulations in this work are performed using the Large-scale Atomic/Molecular Massively Parallel Simulator (LAMMPS) \cite{Thompson2022}. To ensure reproducibility and comparability across workflows, a consistent set of fundamental configuration parameters is used. We use the \texttt{metal} unit system \cite{LAMMPS_units}, which is standard for metallic materials and provides energies in electronvolts, distances in \AA\ and times in picoseconds.

In general, the simulation protocols follow a standardized sequence: velocities are initialized from a Gaussian distribution at the target temperature, the system is equilibrated in the isothermal-isobaric ensemble to remove residual stresses and establish thermal equilibrium, and the production simulation is then performed, during which the relevant physical quantities are recorded. All simulations make use of the domain decomposition and parallelization capabilities provided by LAMMPS, enabling efficient execution of workflows on large systems and on high-performance computing clusters.\\

\subsubsection{Equation of state: equilibrium volume and bulk modulus} \label{workflow:eos}

The equation of state, or energy–volume curve, describes the variation in the total energy of a system as a function of its volume near equilibrium. The equation of state gives access to a number of material properties, such as the equilibrium energy and volume, and the bulk modulus. The first step in this workflow is the creation of atomic structure. We first relax the structure, and then calculate the energies in a given volume range. We calculate the energy of a given input atomic structure at different volumes and fit the resulting data using the Birch–Murnaghan equation of state \cite{Birch1947}.



\subsubsection{Elastic properties: elastic constants and derived moduli} \label{workflow:elastic}

Elastic constants are fundamental properties of crystalline materials and characterize the relationship between stress and strain in the elastic regime. The elastic tensor of an anisotropic material has 21 components represented as a 6-by-6 matrix $C_{ij}$. The bulk modulus $B$ quantifies a material's resistance to homogeneous compression, and the shear moduli $G_1$ and $G_2$ indicate its resistance to shear deformation in distinct crystallographic directions. Poisson's ratio $\nu$ defines the transverse contraction in relation to the longitudinal extension when subjected to uniaxial stress. These quantities can be calculated from the elastic constants following Eq. \ref{eq:elastic_relations}. We first create an atomic structure. Then, we calculate these quantities at 0 K by deforming the simulation box and measuring the stress response, following the approach in LAMMPS \cite{LAMMPS_ELASTIC}.

\begin{equation}
\label{eq:elastic_relations}
B = \frac{C_{11} + 2C_{12}}{3}, \qquad
G_1 = C_{44}, \qquad
G_2 = C' = \frac{C_{11} - C_{12}}{2}, \qquad
\nu = \frac{3B - 2G}{2(3B + G)} \, .
\end{equation}

\subsubsection{Mechanical loading: stress-strain response} \label{workflow:stress}

We provide workflows for two common types of mechanical response tests that can be performed in MD simulations: compression tests and tensile tests. These tests can be applied either uniaxially or hydrostatically and enable the calculation of several mechanical properties, including stiffness, yield stress, flow stress and indicators of phase transitions or phase stability under pressure. In a first step, the atomic structure is created. This structure is then used to created a polycrystalline structure of given grain size using \texttt{atomsk} \cite{Hirel2015}. This polycrystal is first equilibrated at a high temperature and subsequently at the target, lower, temperature to remove initial stresses and achieve a realistic grain boundary structure. Such a sample is then subjected to hydrostatic compression or uniaxial tension along the z axis at a specified strain rate. During the simulation, both the stress response and the deformation of the simulation cell are recorded.

\subsubsection{Thermal properties: thermal expansion and specific heat} \label{workflow:thermal}

Here, we calculate the thermal expansion coefficient $\alpha_P$ and specific heat $c_P$, that describe the coupling between temperature and mechanical response of a system. To this end, an atomic structure is created, which is then equilibrated in the isothermal-isobaric ensemble (NPT). After the equilibration stage, an MD simulation in the same ensemble is performed, and the internal energy at each timestep is recorded. From the fluctuations in this ensemble, the specific heat at constant pressure, $c_P$, can be determined as \cite{Allen2017}

\begin{equation}
\langle (\delta(\mathcal{U} + PV))^2 \rangle_{NPT} = k_B T^2 c_P,
\end{equation}
where $\mathcal{U}$ is the internal energy, and $P$ and $V$ are the instantaneous pressure and volume, respectively, at temperature $T$. $k_B$ denotes the Boltzmann constant. The same fluctuation formalism also provides an estimate of the thermal expansion coefficient, $\alpha_P$

\begin{equation} \label{eq:thermexp}
\langle \delta V \, \delta(\mathcal{U} + PV) \rangle_{NPT} = k_B T^2 V \alpha_P.
\end{equation}

\subsubsection{Free energy calculations: Gibbs free energy and phase stability} \label{workflow:free_energy}

Gibbs free energies govern phase stability under pressure. We provide workflows for the calculation of both Gibbs and Helmholtz free energies. The free energy calculation starts with the creation of an atomic structure, which is then equilibrated at the given temperature and pressure. The free energy of the  crystal at this temperature and pressure is then calculated using non-equilibrium thermodynamic integration methods, followed by the determination of the free energy as a function of pressure. Both methods are employed as implemented in \texttt{calphy} \cite{Menon2021}. 

\subsubsection{Nanoindentation: hardness and load-displacement response} \label{workflow:nano}

Nanoindentation is a widely used technique for investigating hardness, elastic modulus, and conditions for dislocation bursting, and stress-induced phase transformations not just in experiments, but also in MD simulations \cite{HuangZhou2017}. The procedure involves moving a nanoindentor to the surface of the material under test and then indenting it to evaluate the material properties. The nanoindentation simulation workflow is designed to mimic experimental measurement protocols to provide insight into the mechanical response of materials at the atomic level.

The different steps in the workflow include creation of atomic structure, preparing substrates with suitable surface geometry and vacuum layer, with subsequent thermal equilibration to remove any residual stresses, followed by a vertical spherical indentor, that indents the material surface with a constant velocity. This enables recording raw force-depth curves alongside time-stamped atomic configurations at specified intervals, which can then be analyzed to compute fitted curves for the extraction of modulus and hardness, as well as to detect pop-in spots and their magnitudes. 

\subsubsection{Defect energetics: point-defect formation energies} \label{workflow:defect}

One important quantity related to crystallographic defects is the formation energy. Here, we present workflows for calculating the formation energy of atoms acting as point defects in a host metal matrix. The workflow begins with the creation of a pristine atomic structure and the calculation of its total energy. Subsequently, a structure with substitutional or interstitial defects is generated, followed by a second total energy calculation. The defect formation energy is then computed using

\begin{equation}
    E_{f}^{X} = E_{tot}^{nH+1X} - (E_{bulk}^{nH}+E_{ref}^{1X}),
\end{equation}

where $E_{tot}^{nH+1X}$ is the total energy of the configuration containing $n$~ host (H) atoms and a single (X) atom occupying either an interstitial or substitutional site, $E_{bulk}^{nH}$ denotes the total energy of the corresponding bulk host metal supercell with the same number of (H) atoms, and $E_{ref}^{1X}$ represents the reference energy of an (X) atom in a reference configuration.

\section{Application Use Cases}

As described in Section \ref{sec:semantic}, we employ a framework that allows us to produce atomistic simulation data in reproducible workflows while simultaneously annotating the data with ontologies, serializing the metadata as RDF triples. In this section, we showcase the utility of this framework by presenting three use cases: (i) we employ the workflows on a prototypical material system and compare the calculated material properties with existing data; (ii) we show how the workflows can be used to compare materials properties across different models; (iii) we validate structure-property relationships. These represent routine and important tasks in the computational materials science domain. In use cases (ii) and (iii), we demonstrate in particular how the use of ontologies with workflows can formalize knowledge about the physical meaning of the methods and material systems employed to produce the data. We aim to illustrate the added value of producing FAIR data with semantic descriptions for accelerating the research process in materials science.

\subsection{Demonstration on a prototypical material system} \label{sec:workflow_results}

In this use case, we apply our workflows on iron as a prototypical system. Iron provides a well established reference material with multiple stable phases and a broad set of validated interatomic potentials, making it suitable for benchmarking and demonstrating the general applicability of our workflows. The choice of an interatomic potential is critical for accurate representation of material properties. We use an Embedded Atom Method \cite{Daw1984} potential (referred to as EAM01 \cite{EAM01Gunkelmann2012}) throughout this work due to its applicability in predicting the properties of iron under pressure, including pressure-induced phase transformations.\\

\begin{figure}[htbp]
    \centering
    \includegraphics[width=0.95\textwidth]{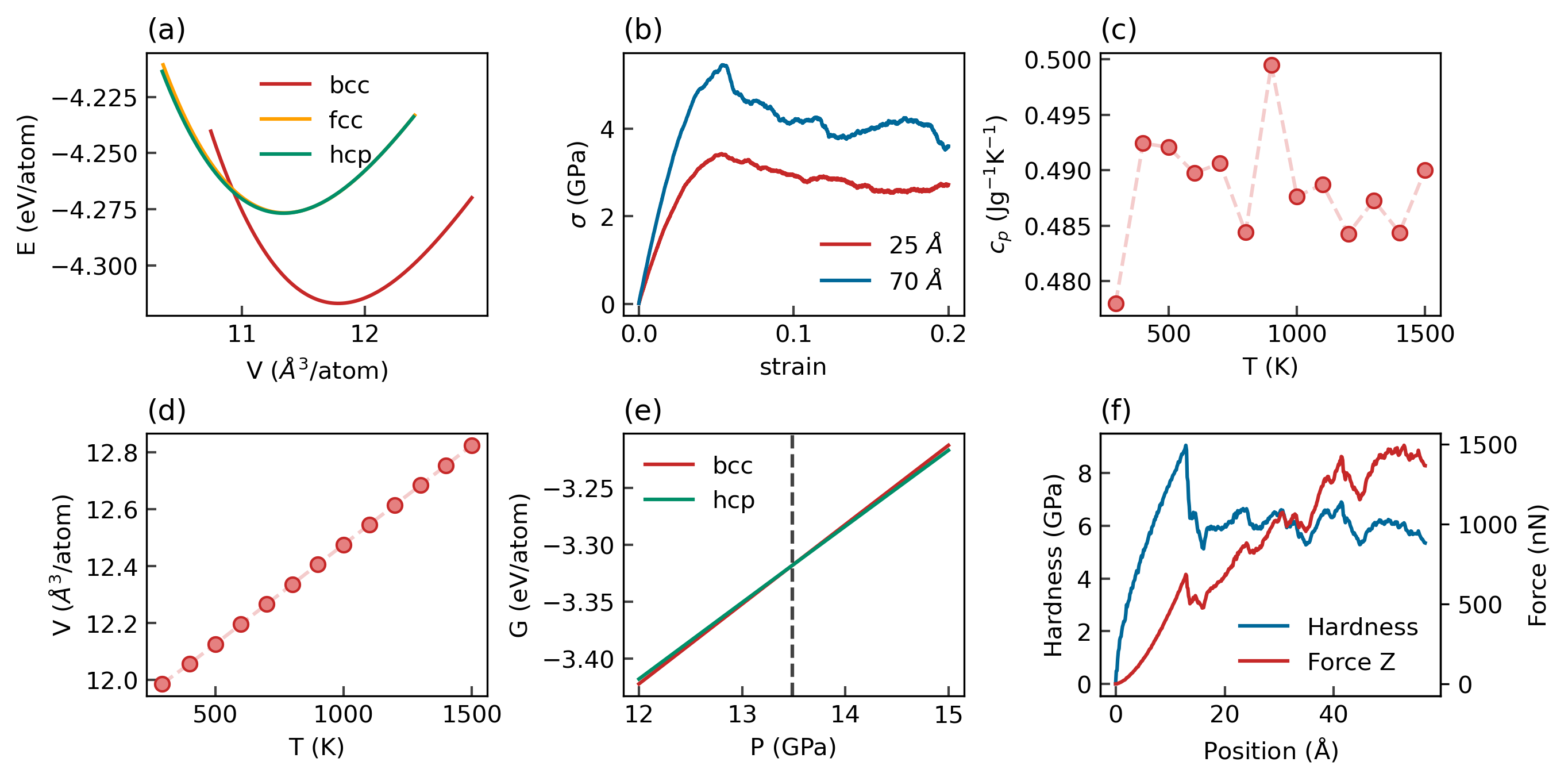}
    \caption{Thermophysical properties of bcc Fe as calculated with the EAM01 potential: (a) Energy-volume curve in comparison with those  of different crystal structures in Fe, (b) flow stress with increasing grain size, (c) specific heat, (d) coefficient of thermal expansion, (e) phase transition bcc to hcp as a function of pressure, (f) hardness and forces from nanoindentation.}
    \label{fig:res-complete}
\end{figure}

\noindent
For the equation of state (Sec. \ref{workflow:eos}), we perform calculations for the body-centered cubic (bcc), face-centered cubic (fcc), and hexagonal close-packed (hcp) crystal structures using the corresponding unit cells. The results of the Birch-Murnaghan equation of state, calculated at 0.5 \% intervals, are shown in Fig. \ref{fig:res-complete} (a). As expected, the bcc structure is identified as the ground state of Fe, while the fcc and hcp structures exhibit very similar energies. The bulk modulus obtained with this potential is 177 GPa, which agrees well with experimental calculated value of 166 GPa \cite{Adams2006}. The bulk modulus can also be calculated from the methods described in Sec. \ref{workflow:elastic}, along with the complete elastic tensor. The tensor elements and derived moduli are summarized in Table~\ref{tab:elastic}. \\

\begin{table}[ht]
    \centering
    \caption{Elastic constants and derived properties for bcc Fe calculated using the EAM potential.}
    \begin{tabular}{lc}
        \toprule
        Quantity & Value (\si{GPa}) \\
        \midrule
        $C_{11}=C_{22}=C_{33}$ & \num{243} \\
        $C_{12}=C_{13}=C_{23}$ & \num{145} \\
        $C_{44}=C_{55}=C_{66}$ & \num{116} \\
        $C'=\tfrac{1}{2}(C_{11}-C_{12})$ & \num{49} \\
        Bulk modulus $B$ & \num{178} \\
        Shear modulus $G_1$ & \num{116} \\
        Shear modulus $G_2$ & \num{49} \\
        Poisson ratio $\nu$ & \num{0.37}$^{*}$ \\
        \bottomrule
    \end{tabular}
    \vspace{0.5ex}
    
    {\footnotesize $^{*}$Poisson ratio is dimensionless.}
    \label{tab:elastic}
\end{table}

\noindent
We employ two workflows for calculating the mechanical response of a material system: stress-strain curves (Sec. \ref{workflow:stress}) and nanoindentation (Sec. \ref{workflow:nano}). We focus on uniaxial tensile tests (Sec. \ref{workflow:stress}) and create polycrystalline bcc Fe structures of specified grain size, equilibrated first at a temperature of 600 K, and then at the target temperature of 10 K. We then employ uniaxial tension along the z axis at a strain rate of $5 \times 10^{-8}\ \mathrm{s^{-1}}$. The polycrystal contains more than ten million atoms, and the simulation is carried out for 300 ps. The resulting stress-strain curves for two grain sizes, 25 and 70 \AA, are shown in Fig.~\ref{fig:res-complete}(b). The curves clearly display the elastic region at low strain, followed by the yield point and the flow stress regime at higher strains. We further perform nanoindentation (Sec. \ref{workflow:nano}) following the simulation procedures and hardness calculations adopted from Luu et al. \cite{luu2021}. In Fig. \ref{fig:res-complete}(f) we show the calculated force and hardness with increasing indentation depth for an Fe system of 54000 atoms.
These examples also demonstrate that our workflow framework supports simulations on extremely large atomistic systems ranging from thousands to millions of atoms. The structured metadata captured at each step ensures that the complete simulation setup, including grain-generation parameters, applied loading conditions, and all thermodynamic details, is stored in the knowledge graph. This data can be queried and analysed, enabling reuse of the generated data.\\

\noindent 
The specific heat and the coefficient of thermal expansion can be calculated using the workflows in Sec. \ref{workflow:thermal}. For these calculations, we use a system consisting of 6750 atoms arranged in a bcc lattice for Fe. The simulations are performed at zero pressure at a range of different temperatures. Each simulation is run for 1~ns, and the instantaneous energy and volume are recorded throughout. The calculated $c_P$ values are shown in Fig. \ref{fig:res-complete} (c). Our results show good agreement with experimental observation of 0.45 $Jg^{-1}K^{-1}$ \cite{Chase1985}. The atomic volume at different temperatures are also obtained, as shown in Fig. \ref{fig:res-complete} (d). The thermal expansion coefficient can be obtained from Eq.~\ref{eq:thermexp} as well as from the slope in Fig.~\ref{fig:res-complete}(d). Using these two approaches, we compute values of $53 \times 10^{-6}\ \mathrm{K^{-1}}$ and $58 \times 10^{-6}\ \mathrm{K^{-1}}$ at 293~K, respectively. Both values exceed the experimentally observed coefficient of $35 \times 10^{-6}\ \mathrm{K^{-1}}$ \cite{Kozlovskii2019}, indicating that the EAM01 potential captures the qualitative temperature dependence but exhibits increased anharmonicity in this regime.\\

\noindent
We calculate the free energy of bcc and hcp crystal structures of Fe in the pressure range of 12--15~GPa using the workflows described in Sec. \ref{workflow:free_energy}. We perform the calculations at 100 K, and use a system size of approximately 2000 atoms for both crystal structure. A phase transition from bcc to hcp is observed at approximately 13.5~GPa. This result is consistent with previous theoretical calculations \cite{EAM01Gunkelmann2012} and experimental observations \cite{Jensen2009}.\\

\noindent
Finally, we use the workflows described in Sec. \ref{workflow:defect} to investigate the energetics of a carbon impurity in iron. The carbon atom occupies octahedral sites as well as substitutional positions in bcc Fe. The calculated energies for this case are reported in Table \ref{tab:e_form}, with simulation cell sizes of $5\times5\times5$, resulting in 250 Fe atoms. To calculate the defect formation energies, we use a universal MLIP, GRACE \cite{GRACElysogorskiy2025}, that can describe both Fe and C. The reference value for the C for the calculation of the formation energy is that of diamond (-9.1 eV/atom). The calculated defect formation energies show good agreement with DFT values, with interstitial and substitutional C lying within the expected energetic range reported in previous studies, such as 0.72 eV for the octahedral C in bcc Fe \cite{HRISTOVA2011}. 

\begin{table}[ht]
    \centering
    \caption{Formation energy of C in bcc Fe with the GRACE potential \cite{GRACElysogorskiy2025}, calculated taking C in diamond as a reference.}
    \label{tab:e_form}
    \begin{tabular}{lc}
		\toprule
		Case & Value (\si{eV})\\
		\midrule
		C octahedral & \num{0.665} \\
		C substitutional & \num{2.935}\\
		\bottomrule
	\end{tabular}
\end{table}

\subsection{Evaluating interatomic potentials through automated workflow execution}\label{sec:pots}

In MD simulations, the calculated properties often exhibit a strong dependence on the chosen model of atomic interactions, that is, the interatomic potential. Therefore, systematic validation tests are necessary to identify and select suitable potentials for a given material or property of interest. 

Here, we demonstrate two key advantages of our workflow framework. First, the workflows are fully reusable and can be easily adapted to different interatomic potentials without modification. Second, because metadata are recorded at every stage of the workflow, all relevant information about the potential, including its type, source, and bibliographic reference, is automatically captured and linked to the generated data. 

As a representative example, we compute the bulk modulus and elastic constants for nine different interatomic potentials. Apart from EAM01, we employ ten other EAM potentials (EAM02 - EAM10) \cite{EAM02Sun2022, EAM03Starikov2021, EAM04Proville2012, EAM05Chiesa2011, EAM06MALERBA201019, EAM07OLSSON2009135, EAM08CHAMATI20061793, EAM09Zhou2004, EAM10Mendelev2003}, as well as an Atomic Cluster Expansion (ACE) MLIP  \cite{Lysogorskiy2021, ACEBienvenu2025}, and a universal Graph Atomic Cluster Expansion (GRACE) potential  \cite{GRACElysogorskiy2025}.

The resulting values are stored in the \texttt{ConceptualDict} object and subsequently parsed into the knowledge graph, enabling structured data storage and semantic querying. This approach allows the data to be queried directly using SPARQL. However, formulating SPARQL queries can be a complex task and often requires detailed knowledge of the underlying ontology \cite{Ngomo2013}. Therefore, we use the \texttt{tools4RDF} library to assist in the generation of queries based on the ontologies in use \cite{tools4RDF}. We nevertheless present the SPARQL queries explicitly to improve clarity and to help understand how the ontology predicates are applied.
Listing~\ref{list:cb2a} shows an example SPARQL query, based on the CMSO and ASMO ontologies, that retrieves both the bulk modulus value and the corresponding potential used in its calculation. The ontology-based structure ensures that all data are stored in a semantically consistent and machine-interpretable manner. 

\begin{listing}[H]
\begin{minted}[linenos, fontsize=\small]{sparql}
PREFIX cmso: <http://purls.helmholtz-metadaten.de/cmso/>
PREFIX rdf: <http://www.w3.org/1999/02/22-rdf-syntax-ns#>
PREFIX asmo: <http://purls.helmholtz-metadaten.de/asmo/>
SELECT DISTINCT ?AtomicScaleSample ?BulkModulusValue ?Reference
WHERE {
    ?AtomicScaleSample asmo:hasCalculatedProperty ?BulkModulus .
    ?BulkModulus asmo:hasValue ?BulkModulusValue .

    ?BulkModulus asmo:wasCalculatedBy ?Simulation .
    ?Simulation asmo:hasInteratomicPotential ?Potential .
    ?Potential cmso:hasReference ?Reference .

   { ?AtomicScaleSample rdf:type cmso:AtomicScaleSample . }
   { ?BulkModulus rdf:type asmo:BulkModulus . }
}
\end{minted}
\caption{SPARQL query that retrieves bulk modulus values together with the associated interatomic potentials and bibliographic references.}
\label{list:cb2a}
\end{listing}

\begin{figure}[htbp]
    \centering
    \includegraphics[width=0.9\textwidth]{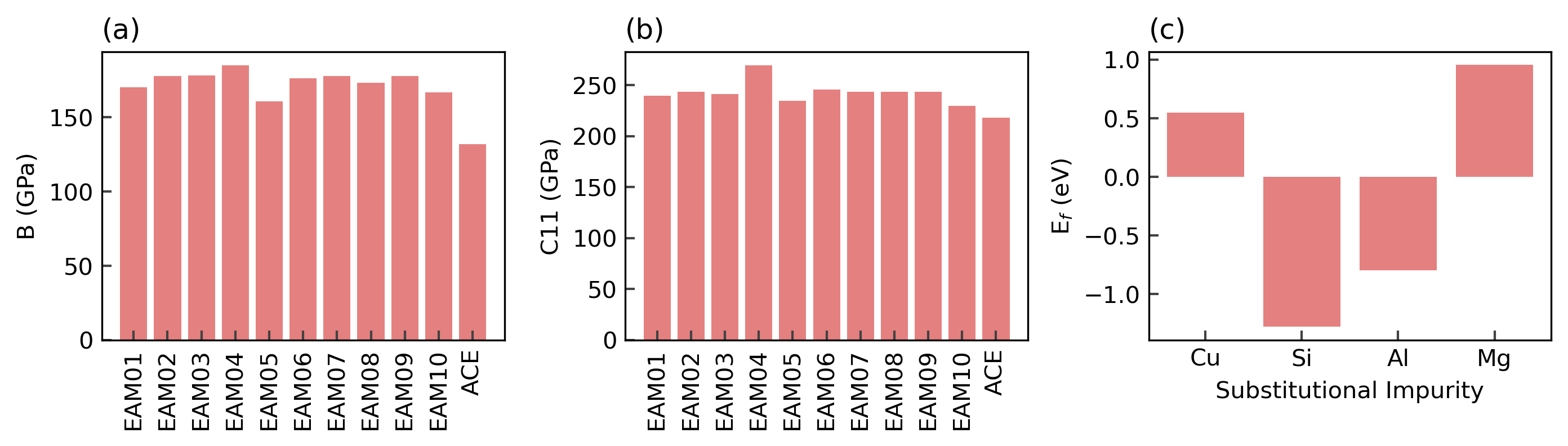}
    \caption{ Results of SPARQL queries from the knowledge graph: (a) Bulk modulus and (b) elastic constant, $c_{11}$, with different EAM potentials. (c) Formation energy of substitutional impurities (X=Cu, Si, Al, Mg) in Fe computed using the GRACE \cite{GRACElysogorskiy2025} model.}
    \label{fig:bc11}
\end{figure}

The software infrastructure we employ, comprising the workflow nodes, the workflow execution environment, the conceptual dictionary, and the knowledge graph creation routines, ensures that identical crystal structures are used across all simulations and that each simulation entry in the knowledge graph retains complete metadata, including details of the interatomic potential and computational parameters. The  bulk modulus values resulting from the SPARQL query in Listing 2 are shown in Fig.~\ref{fig:bc11}(a).

A similar SPARQL query for the elastic constant $C_{11}$ provides the corresponding extracted values, presented in Fig.~\ref{fig:bc11}(b). This demonstrates how specific material properties can be programmatically retrieved from the knowledge graph, along with their complete provenance information. Such ontology-based data representation enhances interoperability across different simulation workflows and facilitates the reuse of both data and metadata within the broader materials science research ecosystem.

We show another example for querying defect formation energies computed using the GRACE \cite{GRACElysogorskiy2025} model. The corresponding SPARQL query is given in Listing \ref{list:cb2c}. This example highlights functionality that becomes available when using our knowledge-based workflows. When a substitutional defect is introduced by replacing an atom, the workflow automatically annotates that the resulting sample contains a substitutional defect. As a result, SPARQL can be used to retrieve all samples that include substitutional impurities. In traditional workflows, where only atomic positions are stored, such information is easily lost and identifying defects could require manual inspection. The SPARQL query returns the defect formation energies together with the chemical species present in the system. The resulting values are shown in Fig. \ref{fig:bc11} (c). With the exception of Si in an Fe matrix, all formation energies fall within the expected 0.1 to 0.4 eV range \cite{GRACElysogorskiy2025} from DFT calculations \cite{Jelinek2012}. These examples collectively demonstrate how our ontology-aware workflows enable systematic comparison of interatomic potentials and provide semantically enriched data suitable for large-scale analysis.

\begin{listing}[H]
\begin{minted}[linenos, fontsize=\small]{sparql}
PREFIX cmso: <http://purls.helmholtz-metadaten.de/cmso/>
PREFIX rdf: <http://www.w3.org/1999/02/22-rdf-syntax-ns#>
PREFIX asmo: <http://purls.helmholtz-metadaten.de/asmo/>
PREFIX podo: <http://purls.helmholtz-metadaten.de/cdos/podo/>
PREFIX cdco: <http://purls.helmholtz-metadaten.de/cdos/cdco/>
SELECT DISTINCT ?AtomicScaleSample ?value ?label ?elements
        (GROUP_CONCAT(DISTINCT ?symbol; separator=", ") AS ?elements)
WHERE {
    ?AtomicScaleSample asmo:hasCalculatedProperty ?property .
    ?AtomicScaleSample cmso:hasMaterial ?material .
    ?material cdco:hasCrystallographicDefect ?defect .
    ?AtomicScaleSample cmso:hasSpecies ?species .
    ?species cmso:hasElement ?element .
    ?element cmso:hasChemicalSymbol ?symbol .
    ?property rdfs:label ?label .
    ?property asmo:hasValue ?value .

    { ?Defect rdf:type podo:SubstitutionalImpurity . }
    FILTER (?label='DefectFormationEnergy')
}
GROUP BY ?AtomicScaleSample ?value ?label
}
\end{minted}
\caption{SPARQL query that retrieves defect formation energies for samples containing substitutional impurities. The query identifies atomic-scale samples, their calculated defect formation energies, and the chemical species present}
\label{list:cb2c}
\end{listing}

\subsection{Querying structure-property relations}

One of the fundamental principles of materials science is to understand structure-property relationships and employ these effectively for materials design. In that aim, the microstructure of the material plays a fundamental role, the presence of crystallographic defects directly affect the physical properties of the material. We propose that semantically annotated data can be leveraged to express such relationships and allow direct querying to validate fundamental physics principles. 

Here, we present the example of the Hall-Petch effect, which gives the relation between the grain size and the strength of metallic materials and is described by:

\begin{equation}
    \sigma_y = \sigma_i + kd^{-1/2}
\end{equation}
where $\sigma_{y}$ is the yield stress (or flow stress at higher strains); $\sigma _{i}$ the stress required to move dislocations through a single crystal or very large grain; $k$ the Hall-Petch coefficient, a constant specific to the material and $d$ is the average grain diameter. The validation of Hall-Petch laws is highly relevant for engineering applications, and similar data-driven analyses have been demonstrated by automatically extracting grain size and yield strength data from the literature \cite{Kumar2022}.

From the data generated using our uniaxial tensile test workflow, described in Sec. \ref{workflow:stress}, for polycrystalline Fe using EAM01 potential, we query computational samples with different grain sizes and the corresponding calculated flow stress, as shown in Listing \ref{list:cb3}. 

\begin{listing}[H]
\begin{minted}[linenos, fontsize=\small]{sparql}
PREFIX cmso: <http://purls.helmholtz-metadaten.de/cmso/>
PREFIX rdf: <http://www.w3.org/1999/02/22-rdf-syntax-ns#>
PREFIX asmo: <http://purls.helmholtz-metadaten.de/asmo/>
SELECT DISTINCT ?AtomicScaleSample ?FlowStress ?FlowStressValue ?Unit ?GrainSize
WHERE {
    ?AtomicScaleSample cmso:hasSimulationCell ?SimulationCell .
    ?SimulationCell cmso:hasGrainSize ?GrainSize .
    ?AtomicScaleSample asmo:hasCalculatedProperty ?FlowStress .
    ?FlowStress asmo:hasValue ?FlowStressValue .
    ?FlowStress asmo:hasUnit ?Unit .
   { ?AtomicScaleSample rdf:type cmso:AtomicScaleSample . }
   { ?FlowStress rdf:type asmo:FlowStress . }
}
\end{minted}
\caption{SPARQL query to extract average grain size and flow stress from the knowledge graph.}
\label{list:cb3}
\end{listing}

\begin{figure}[htbp]
    \centering
    \includegraphics[width=0.45\textwidth]{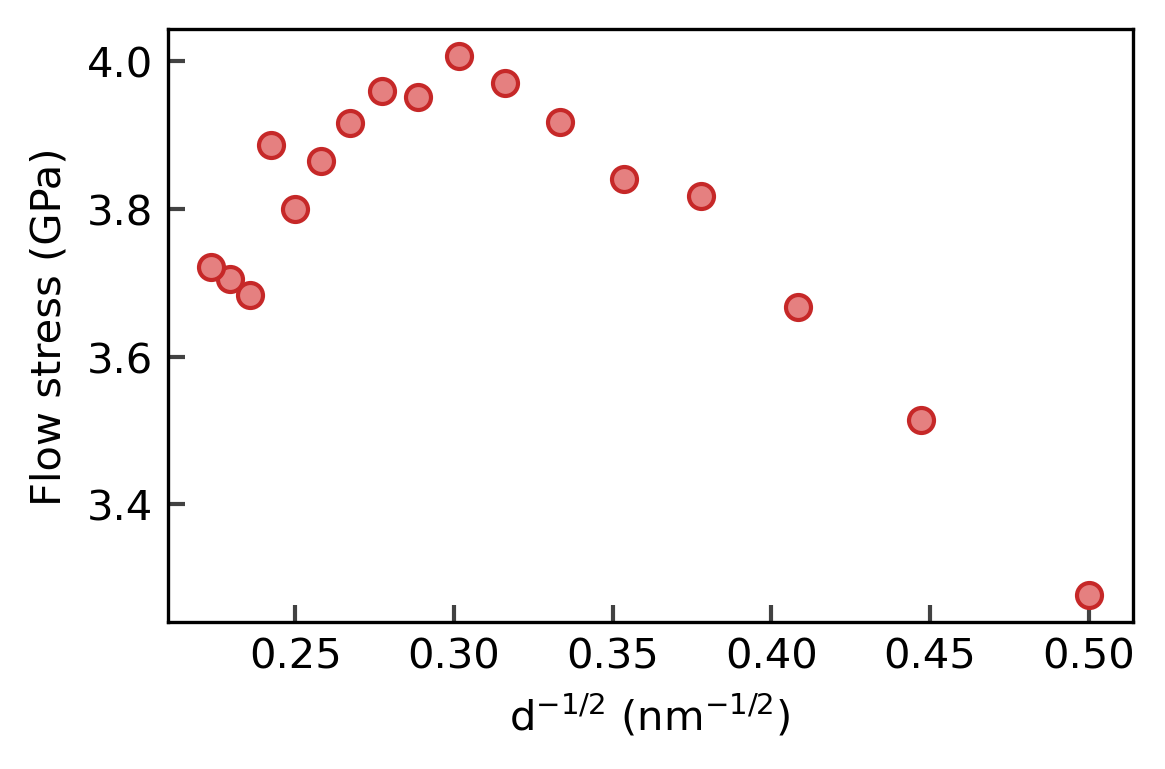}
    \caption{Variation of flow stress with inverse square root of the average grain size. }
    \label{fig:hallpetch}
\end{figure}

The resulting values are then plotted in Figure \ref{fig:hallpetch}, which depicts the average flow stress as a function of the inverse square root of the average grain size, both quantities are automatically annotated at the workflow stage. We see the grain boundary strengthening mechanism according to the Hall-Petch equation, initially the stress increases with decreasing grain size, until the strengthening limit after which the stress is reduced due to grain boundary sliding. This softening is known as the inverse Hall-Petch effect.

\section{Discussion}

This work illustrates that combining workflow orchestration with semantic metadata enables a knowledge-based approach that significantly enhances the interoperability and interpretability of atomistic simulation data. We provide workflows for mechanical and thermodynamic properties calculations and produce annotated datasets specific to the Fe and Fe-X material systems.

The use cases demonstrate that semantically annotated workflows can be used for data-driven validation of physical laws, such as the Hall–Petch effect, directly from existing simulation data. The framework presented in this work also supports systematic comparison across interatomic potentials, which is essential for uncertainty quantification in atomistic modeling. The workflows intentionally capture more complex systems and simulations such as tensile and compression tests, free energy calculations, nanoindentation and defect formation energies, showing that semantic workflow design is scalable to mechanical behavior and phase transformation scenarios that are relevant for engineering applications.

\begin{figure}[htbp]
    \centering
    \includegraphics[width=0.4\textwidth]{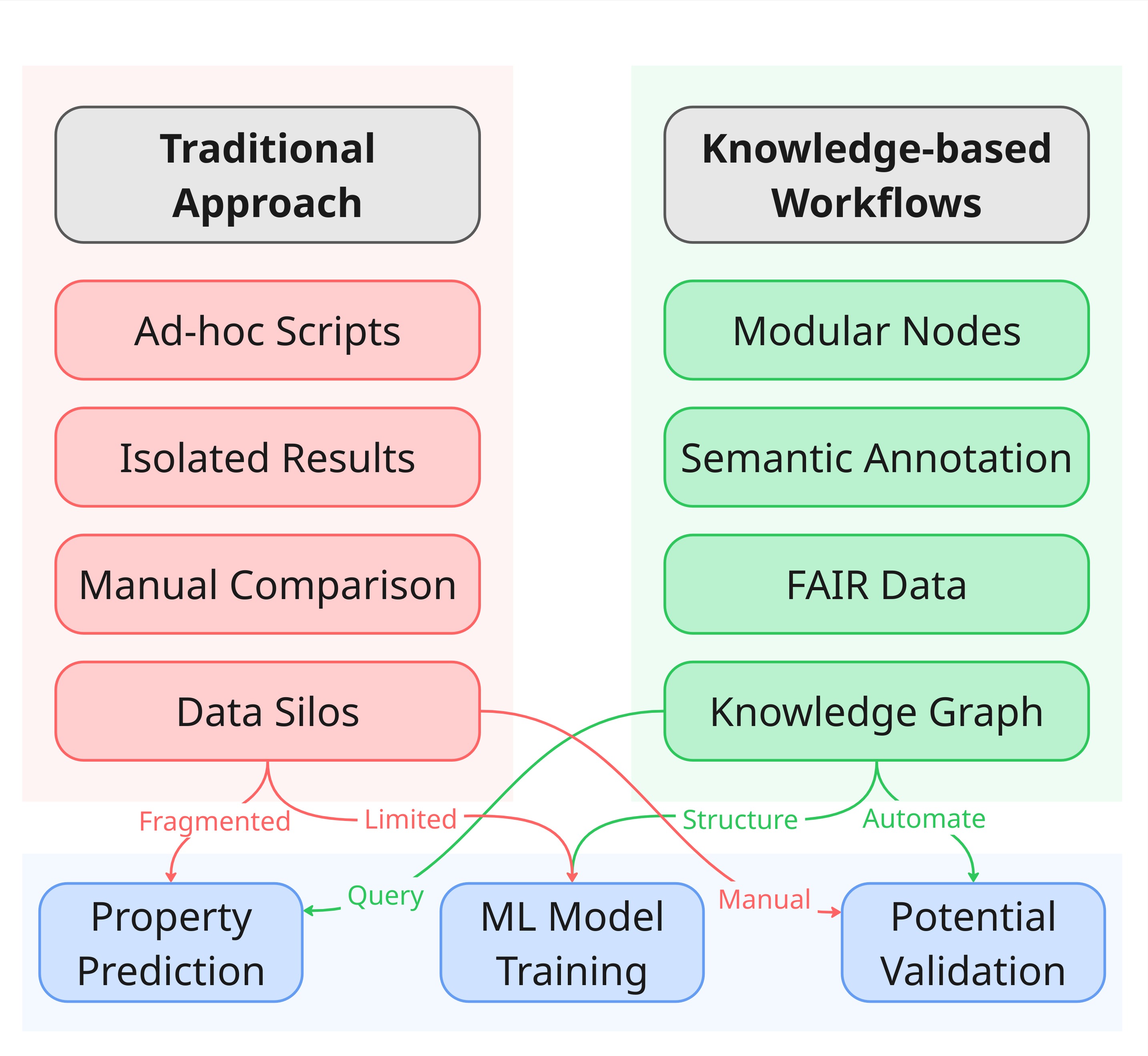}
    \caption{Comparison of traditional ad-hoc simulation scripts (left) and knowledge-based workflows (right). Traditional approaches produce isolated results that require manual comparison and lead to data silos. Semantic workflows generate FAIR-compliant data and queryable knowledge graphs that improve interoperability and reuse.}
    \label{fig:workflow_comparison}
\end{figure}

Figure \ref{fig:workflow_comparison} indicates that knowledge-based workflows overcome the limitations of traditional ad-hoc simulation practices by introducing modular workflow nodes, semantic annotation, and FAIR data.

The added value of the proposed infrastructure can be best illustrated considering the implications for interoperability and reusability:

\begin{enumerate}[label=\roman*.]
    \item \textit{Interoperability:} in terms of FAIR data, the captured metadata is aligned with application and domain level ontologies, providing a common vocabulary to describe the material system, computational sample, simulation workflow and calculated properties. The use of consistent semantic descriptors enable linking outputs from DFT calculations data with classical MD and MLIP simulations in the same knowledge space. This framework also supports semantic integration with external resources, such as materials databases, and has the potential to interoperate with experimental data. In terms of FAIR workflows, our nodes are agnostic to the specific workflow environment. This is demonstrated in Listings~\ref{list:cb1} and \ref{list:cb1job}, where two different workflow systems, pyiron and jobflow \cite{Rosen2024}, are used. Further connections to other workflow systems are possible, for example using the Python Workflow Definition \cite{Janssen2025}. Overall, this shows that the ontology-aware workflow nodes can be seamlessly reused across different workflow systems.
    
    \item \textit{Reusability:} the central advantage of the proposed approach for generating simulation workflows and data is the reusability for different tasks. Workflows can be executed with different interatomic potentials without modifying the code, as demonstrated by the systematic evaluation of 10 different potentials in section \ref{sec:pots}. The semantic layer captures both the physical concepts and the procedural steps of each simulation, allowing the same workflow to be reused across different material systems or thermodynamic conditions by adjusting only the relevant metadata parameters. This capability significantly increase the automation potential for generating comparable datasets for benchmarking or model validation and supports systematic studies of methodological uncertainty in atomistic simulations.
\end{enumerate}

Although the proposed framework improves interoperability and reusability in the production of FAIR data and workflows, there are several limitations that must be acknowledged:

\begin{itemize}
    \item Workflows for calculation of properties relevant for engineering applications remain strongly application-specific and require domain expertise for correct configuration and interpretation. Molecular dynamics is fundamentally constrained by accessible time scales, which constrains realistic loading rates and transformation kinetics. Consequently, the selection of appropriate input parameters still depends on knowledgeable users, and poor choices directly affect the quality of the results. In addition, sensitivity to boundary conditions, such as cell shape, loading mode, or thermostat and barostat algorithms, continues to affect the comparability of simulation outcomes. The semantic layer can document these choices, but it cannot yet eliminate the underlying physical dependencies.
    \item The generalizability of workflows to entirely new simulation scenarios is not automatic. Accurate semantic annotation requires domain expertise, and both the ontologies and the workflow templates need to be extended to new physical processes or modelling methods. Nevertheless, the general architecture is expandable to new simulation scenarios. Other approach is that of \texttt{semantikon} \cite{semantikon}, which offers greater flexibility by allowing serialization with any ontology, although shifts the responsibility of ensuring consistent and meaningful semantic annotation to the user.
    \item Finally, with respect to integration into broader semantic infrastructures, our highest-level alignment currently relies on PROV-O. While PROV-O provides lightweight and flexible provenance representation, alignment with upper-level ontologies such as the Basic Formal Ontology (BFO) \cite{Otte2022} would further enhance interoperability. Existing mappings of PROV-O to BFO indicate that the two are not incompatible \cite{Prudhomme2025}, but adopting a top-level-aligned design would improve integration within the wider materials science semantic ecosystem.
\end{itemize}

Future directions include deeper integration with external knowledge graphs broader research data infrastructures to enable federated search across simulation results, experimental measurements, and literature-derived knowledge. Another opportunity is connecting the semantic workflow framework to emerging agentic AI pipelines that can autonomously select workflows, choose interatomic potentials, propose validation steps, and generate new simulation data. An example of such is \texttt{LangSim}, a library that couples large language models with simulation workflows \cite{zimmermann2025}. The semantic layer can enhance transparency and trustworthiness in such AI-driven approaches. In addition, expanding the underlying ontologies to capture microstructure evolution, more complex defect interactions, and advanced thermodynamic descriptors will allow the framework to support a broader range of physical phenomena and materials modelling tasks. Ultimately, semantically annotated workflows not only serve as documentation tools but as a framework to accelerate understanding and generation of insights from data, validation of models and scientific reproducibility. 

\section{Conclusion and Outlook}

In this work we introduced knowledge-based workflows for atomistic simulations of mechanical and thermodynamic properties, demonstrating how metadata annotation, ontology alignment, and provenance tracking transform traditional simulation scripts into structured and interoperable research objects. By decomposing simulations into modular workflow components bound to domain and application ontologies, we enabled consistent description of materials, computational samples and methods, and calculated properties across a broad set of atomistic simulation tasks. The workflows developed for Fe use cases illustrate how this approach improves reusability, facilitates systematic comparison across datasets, and supports transparent interpretation of material properties.

A central outcome of this work is enhanced interoperability and reusability. The semantic layer harmonizes metadata across simulation software and modelling approaches, consistently annotating data from workflows used with different materials systems, potentials, or boundary conditions. This interoperability extends naturally to institutional databases and national research data infrastructures solutions within NFDI-MatWerk, enabling integration with experimental datasets, external simulation repositories, and emerging knowledge graph ecosystems.

The structured metadata and modular workflow design also position the framework for future AI-driven materials discovery. Semantically annotated datasets support targeted query of physical concepts and validation of structure–property relationships, while agentic AI systems can leverage reusable workflow nodes to select simulation protocols, propose validation tasks, and orchestrate large-scale data generation. Such capabilities provide a pathway toward automated exploration of mechanical properties, high-throughput screening of phase stability, and informed alloy design workflows.

Looking ahead, the framework can be extended to more complex simulation scenarios, e.g. diffusion processes and defect-defect interactions, and multiscale coupling with mesoscale or continuum models. As the library of reusable workflows and materials science ontologies grows, the field can shift from script-based simulations to knowledge-driven computational materials science. We therefore encourage the community to adopt, refine, and expand the workflows and ontologies introduced here, advancing a shared semantic foundation that supports reproducible, scalable, and AI-ready computational materials science.

\medskip
\textbf{Data Availability Statement}  \par
The workflows are provided here \url{https://github.com/pyscal/semantic-workflows}. The data and corresponding metadata is available here \url{https://doi.org/10.5281/zenodo.18380870}.

\medskip
\textbf{Acknowledgements} \par 
This work is supported by the consortium NFDI-MatWerk, funded by the Deutsche Forschungsgemeinschaft (DFG, German Research Foundation) under the National Research Data Infrastructure – NFDI 38/1 – project number 460247524. H.-T.L. and N.M. acknowledge funding by the DFG (Project-ID 394563137 – SFB 1368 and ME 6073/20-1). A.A.G. gratefully acknowledges computing time on the supercomputer JURECA \cite{JURECA} at Forschungszentrum Jülich under grant project ID {\tt defectskg}.

\medskip

\bibliographystyle{MSP}
\bibliography{references}

\end{document}